\documentclass[aps,epsfig,twocolumn]{revtex4}
\def\be{\begin{eqnarray}}
\def\ee{\end{eqnarray}}

\newcommand{\bp}{{\bf p}}

\usepackage{bm}
\usepackage{psfig}
\usepackage[dvips]{graphicx}

\begin{document}
\title{Spontaneous breaking of rotational symmetry 
in superconductors\footnote{Phys. Rev. Lett. 88, 252503 (2002)}}
\author{H. M\"uther and A. Sedrakian}
\affiliation{Institut f\"ur Theoretische Physik, Universit\"at
T\"ubingen, D-72076 T\"ubingen, Germany}

\begin{abstract}
We show that homogeneous superconductors with broken 
spin/isospin symmetry lower their energy via a transition  
to a novel superconducting state where
the Fermi-surfaces are deformed to a quasi-ellipsoidal form
at zero total momentum of Cooper pairs. In this state,
the gain in the condensation energy of the  pairs
dominates over the loss in the kinetic energy caused by the 
lowest order (quadrupole) deformation of Fermi-surfaces from 
the spherically symmetric form. There are two energy minima 
in general,  corresponding to the  deformations of the Fermi-spheres 
into either prolate or oblate forms. The phase transition from 
spherically symmetric state to the superconducting state with broken
rotational symmetry is of the first order.  
\end{abstract}

\maketitle 


In the original Bardeen-Cooper-Schrieffer (BCS) theory of bulk
superconductivity, the condensate wave-function describes a 
quantum coherent state which is invariant under spatial and time
(particle-hole) reversal transformations~\cite{BCS}.
External perturbations which act on discrete quantum variables, like the 
spin of the fermions, break the particle-hole
symmetry. A typical example is a metallic superconductor in a high
magnetic field where the Pauli paramagnetism induces an asymmetry in
the populations of the spin-up and spin-down electrons. The
superconducting state is quenched via a first order phase transition 
once the splitting in the energy spectrum of spin-up and spin-down 
electrons becomes of the order of the pairing gap in the unpolarized
state ~\cite{METALS}.
The crossover from the BCS to the normal state can be understood
in terms of the phases space overlap between the fermionic states 
located at the top of their individual Fermi-surfaces.
The pairing gap is maximal for the symmetric (unpolarized) 
state with perfectly overlapping Fermi-surfaces. As these are driven 
apart by the ``polarizing field'' the phase-space available 
for the pairing decreases and the gap is successively suppressed.
At finite temperatures the smearing of the Fermi-surfaces 
increases the phase-space overlap and hence the critical field
at which the superconductivity is quenched. 

The superconducting state sustains larger asymmetries if
the translational symmetry is broken. Larkin and Ovchinnikov 
and Fulde and Ferrell (LOFF) argued that the crossover from 
the BCS to the normal state, as the Fermi-surfaces are 
driven apart by the polarizing field, occurs via  a spatially inhomogeneous 
superconducting phase~\cite{LARKIN,FULDE}. The Cooper pairs carry 
a finite total momentum in the LOFF phase, i.e., the centers of the
Fermi-spheres are shifted allowing for a partial phase-space overlap.  

This paper suggests an alternative mechanism of breaking  
the symmetry which is based on a deformation of the spherical
Fermi-surfaces, to the lowest order, into quasi-ellipsoidal form. 
If the total momentum of the Cooper pairs is zero, as we shall 
assume in the following, the deformation spontaneously breaks the 
rotational symmetry. The novel superconducting phase 
maintains its stability due to the dominance of 
the condensation energy of the Cooper pairs over 
the loss in the kinetic energy of the system caused by the 
deformation of the Fermi-surfaces. We shall assume that there
is a single axis along which the symmetry is broken, although more 
complicated patterns of symmetry breaking are possible
(simultaneous breaking of the rotational and translational 
symmetries, higher order multipole deformations of Fermi-surfaces, etc). 
Note that deformed or non-spherical Fermi-surfaces are common for electrons
in solids; here we treat systems which are  homogeneous in the 
normal state, i.e., any deformations of the spherical
shape of the Fermi-surfaces would correspond to a metastable state.

We start with the BCS-Gorkov equations in energy-momentum representation
\be\label{QC}
\sum_{\gamma}\left( \begin{array}{cc}
\omega - E_{\alpha\gamma}
       &-\Delta_{\alpha\gamma}\\
       - \Delta_{\alpha\gamma} ^{\dagger}
& \omega + E_{\alpha\gamma}
      \end{array} \right) \left( \begin{array}{cc}
        G_{\gamma\beta} & F_{\gamma\beta}\\
        F_{\gamma\beta}^{\dagger}
        & \overline G_{\gamma\beta}
      \end{array} \right)= \delta_{\alpha\beta} \hat {\bf 1},
\ee
where $G_{\gamma\beta}(\omega,\bp)$ and $F_{\gamma\beta}(\omega,\bp)$ 
refer to the full normal and anomalous retarded propagators,
$\Delta_{\alpha\gamma}(\omega,\bp)$ is the anomalous self-energy,  
and the diagonal matrix elements of the first matrix correspond to the 
inverse of free-single particle propagators;
the Greek indeces $\alpha, \beta \dots = 1,2$ label the 
two different species, and $\omega$ and $\bp$ refer to the 
particle energy and momentum. (Note that the center-of-mass 
momentum of particles is zero). 
Suppose that the rotational symmetry is broken by 
a deformation of the Fermi-surfaces from spherical form. 
The quasiparticle spectrum of species $\alpha$ 
can be parameterized, then, as
\be
\label{SPEC1}
E_{\alpha}= \frac{p^2}{2m_{\alpha}}
-\mu_{\alpha}\left(1-\epsilon_{\alpha}\cos^2\theta\right),
\ee
where $\mu_{\alpha}$ are the chemical potentials of the particles in
the undeformed state, $\theta$ is the angle between 
the particle momentum $p$ and the axis of symmetry breaking; 
the deformation of the Fermi-sphere in  Eq. (\ref{SPEC1}) 
is truncated at the lowest order non-trivial axisymmetric
deformation, which is described by the $l = 2$, $m=0$ term of  
the Legendre polynomials associated with an expansion in spherical
harmonics. The constant energy surfaces of quasiparticle excitations 
defined by Eq. (\ref{SPEC1}) represent quasi-ellipsoids of 
revolution with an ellipticity $\epsilon_{\alpha}$. 
For $\epsilon_{\alpha} = 0$ the spectrum (\ref{SPEC1}) is the 
true eigenstate of the unpaired, homogeneous system in the absence of external 
fields. (Note that the deformation described by
Eq. (\ref{SPEC1}) does not need to conserve the volume of the
Fermi-sphere, as we impose a self-consistency condition 
for the total density of the system, see Eq. (\ref{DENSITY}) below).

In the following we shall neglect the possible pairing among the same 
species ($\Delta_{\alpha\alpha}=0$) so that only the
off-diagonal elements of the anomalous self-energy matrix are non-zero.
The quasiparticle excitation spectrum in the superconducting phase 
is determined in the standard fashion by finding the poles of the 
propagators
in Eq. (\ref{QC}):
\be\label{SPECTRUM}
\omega_{1,2} =E_A \pm \sqrt{E_S^2+\vert \Delta\vert^2},
\ee
where the symmetric and anti-symmetric
parts of the spectrum (which are even and odd
with respect to the time-reversal symmetry) are defined as $E_{S,A}=
(E_1\pm E_2)/2$.
The solution of Eq. (\ref{QC}) can be written in 
terms of the eigenstates (\ref{SPECTRUM}) as
\be
G_{1,2} &=& \frac{u_p^2}{\omega-\omega_{1,2}+i\eta}
  +  \frac{v_p^2}{\omega-\omega_{2,1}+i\eta}, \\
F &=& u_p v_p \left( \frac{1}{\omega-\omega_{1}+i\eta}
      -\frac{1}{\omega-\omega_{2}+i\eta}\right),
\ee
where the Bogolyubov amplitudes are
\be
u_p^2 = \frac{1}{2} + \frac{E_S}
          {2\sqrt{E_S^2+\vert \Delta\vert^2}} , \quad
v_p^2 = \frac{1}{2} - \frac{E_S}
          {2\sqrt{E_S^2+\vert \Delta\vert^2}} .
\ee 
The mean-field approximation to the anomalous self-energy yields
the gap equation
\be\label{GAP2}
\Delta (\bp) = 2\int \frac{d\omega' d\bp'}{(2\pi)^4}
V(\bp,\bp'){\rm Im} F (\omega',\bp') f(\omega'),
\ee
where $V(\bp,\bp')$ is the bare interaction~\footnote{For the 
sake of simplicity we ignore the renormalization of the 
pairing interaction in the particle-hole channel and 
assume a time-local interaction.},
$f(\omega)=\left[{\rm exp}(\beta\omega)+1\right]^{-1}$is the Fermi
distribution function and $\beta$ is the inverse temperature.
The $\omega$ integration is straightforward in the quasiparticle
approximation, since the frequency dependence of the propagator
is constrained by the on-shell condition. For $S$-wave
interactions the potential depends only on the absolute magnitude 
of the quasiparticle momenta.
In this case the gap equation simplifies to 
\be\label{GAP3}
\Delta( p) &=&
\int\frac{p^{'2}dp'}{(2\pi)^2}
V(p, p')\int d\cos\theta'\nonumber\\&\times &
\frac{\Delta( p')}{2\sqrt{E_S^2+ \Delta(p')^2}}
\left[f(\omega_1)-f(\omega_2)\right].
\ee
Note that the deformation of the Fermi-surfaces enters the gap equation 
as a parameter which is determined by the minimum of the ground state 
energy of the superconducting phase. For strongly coupled superconductors 
the gap equation (\ref{GAP3}) is supplemented by the 
normalization condition for the net density $\rho\equiv \rho_1+\rho_2$
at a constant temperature. The densities of the species are given by
\be\label{DENSITY}
\rho_{1,2} &=& 
-2\sum\int \frac{d^4 p}{(2\pi)^4}{\rm Im}
G_{1,2}(\omega,\bp) f(\omega)\nonumber\\
&=&
\sum\int\frac{d^3p}{(2\pi)^3}
\left\{u_p^2 f(\omega_{1,2})+v_p^2 f(\omega_{2,1})\right\},
\ee
where the summation is over the discrete quantum variables. The second 
equality follows in the quasiparticle approximation.

Next we turn to the thermodynamic properties of the superconducting state.
At a fixed density and temperature the relevant thermodynamic
potential is the free energy:
\be\label{GR}
F\vert_{\rho,\beta} =  U - \beta^{-1}S,
\ee
where $U$ is the internal energy and $S$ is the entropy.
In the mean-field approximation the entropy is given by 
the expression
\be\label{ENTROPY}
S &=& 
- k_B\sum \int\frac{d^3p}{(2\pi)^3}
\Bigl\{f(\omega_{1})\,{\rm ln}\, f(\omega_{1})+
        \bar f(\omega_{1})\,{\rm ln}\,\bar f(\omega_{1})
\nonumber\\&+&
f(\omega_{2})\,{\rm ln}\, f(\omega_{2})
+\bar f(\omega_{2})\,{\rm ln}\, \bar f(\omega_{2})
\Bigr\},
\ee
where $\bar f(\omega_{\pm}) = [1-f(\omega_{\pm})]$, $k_B$ is the
Boltzmann constant. The internal energy, defined as the grand canonical
statistical average of the Hamiltonian, is 
\be\label{U}
U &=& 
\sum\int \frac{d^3p}{(2\pi)^3}\Biggl\{
\left[n_{1}( p)
E_1(p)+n_{2}( p)E_2(p) \right]
\nonumber\\&+&
\int\frac{d^3p'}{(2\pi)^3}
\, V( p , p')\, 
\nu( p)\nu( p')\Biggl\},
\ee
where
\be
n_{1,2}( p)&\equiv & u_p^2 f(\omega_{1,2})+v_p^2 f(\omega_{2,1}) ,
\\
\nu( p ) &\equiv & u_pv_p\left[f(\omega_1)-f(\omega_2)\right].
\ee
The first term in Eq. (\ref{U}) is the kinetic energy while the 
second term includes the mean field interaction among 
the particles in the condensate.
The true ground state of the system minimizes the free energy
difference  $\delta F\vert_{\rho,\beta}$ between the superconducting
and normal states (the free energy in the normal state
follows from Eqs. (\ref{ENTROPY}) and
(\ref{U}) when $\Delta= 0$).

The deformations of the Fermi-spheres can be described 
in terms of the ``conformal deformation'' $\epsilon =
(\epsilon_1+\epsilon_2)/2$ and  the ``relative deformation''
$\delta\epsilon = (\epsilon_1-\epsilon_2)/2$. 
Then, the symmetric and anti-symmetric parts of the energy spectrum
can be written as 
\be
E_S &\equiv & \frac{p^2}{2m}-\mu\left[1+\epsilon\cos^2\theta
\left(1+\frac{\delta\epsilon\delta\mu}{\epsilon\mu}\right)\right],
\\ 
E_A &\equiv & -\delta\mu
+\left(\mu\delta\epsilon+\epsilon\delta\mu\right)\cos^2\theta ,
\ee
where $\mu =( \mu_1+\mu_2)/2$, $\delta\mu = (\mu_1-\mu_2)/2$ (here we
ignore the difference in the masses of the spin/isospin up and down 
quasi-particles).
Equations (\ref{GAP3}), (\ref{DENSITY}) and (\ref{GR}) form a closed
system, which determines the pairing gap and the ground state energy 
of  a superconductor for constant density asymmetry
$\alpha = (\rho_1-\rho_2)/(\rho_1+\rho_2)$. The values of the 
deformation parameters $\delta\epsilon$ and $\epsilon$ are obtained 
by requiring that the free-energy attains its minimum. Note that 
in the weak coupling limit Eqs. (\ref{GAP3}) and (\ref{DENSITY})
decouple and one may solve for $\Delta$ as a function of $\delta\mu$
instead of $\alpha$. Apart from the fact that $\alpha$, rather than 
$\delta\mu$, is the measurable quantity, there is an additional reason
for solving the full set of equations. The gap equation alone is symmetric
under exchange $E_A \to -E_A$, which implies that 
the solutions are symmetric under the simultaneous change of the 
signs of $\delta\mu$ and $\delta \epsilon$. Eq. (\ref{DENSITY}), 
however, does not have this symmetry and the solutions are distinct
under the sign transformation above. 

As a specific example, which illustrates the solutions above, we
consider isospin-singlet (neutron-proton) pairing in nuclear matter
in the $^3S_1$-$^3D_1$ channel~\cite{SD_PAIRING1,SD_PAIRING2,SD_PAIRING3}.
The gap in the isospin symmetric case is 
$\Delta_{00} = 12$ MeV if we use as the bare interaction the 
Argonne potential and ignore the renormalization of the mass of the
particles in the normal state due to interactions. 
The modification of the particle self-energy in nuclear medium
(see for a review ~\cite{MUETHER}) affect the absolute magnitude of the 
gap and rescale its dependence on the parameters.
Clearly, with these approximations,  
our model is schematic, however we do not expect qualitative changes
when renormalization of the interaction and the bare mass are included.

The BCS solutions for the $n$-$p$ pairing has been studied for the 
homogeneous (translationally and rotationally) invariant state under 
isospin asymmetric 
conditions~\cite{NUCL_BCS1,NUCL_BCS2,NUCL_BCS3,NUCL_BCS4} and the 
inhomogeneous state with broken translational symmetry
~\cite{NUCL_LOFF,COMMENT} (the nuclear analog of the LOFF phase.
(The flavor asymmetric $<qq>$-condensate in high density 
QCD is another example of strongly coupled superconductor where 
the breaking of translational symmetry plays a role 
~\cite{KRISHNA}).
Consistent with the assumption $\Delta_{\alpha\alpha} = 0$ above we 
ignore the $n$-$p$ pairing in the $^1S_0$ channel ~\cite{MORTEN}.

Fig. 1 shows the pairing gap as a function of the density asymmetry 
and the relative deformation $\delta\epsilon$ for vanishing conformal 
deformation ($\epsilon = 0$). Here and below we assume a density 
$\rho = 0.16 $ fm$^{-3}$ and a temperature $T = 3$ MeV. 
The gap parameter is normalized to its value  $\Delta_{00}$ in the
isospin symmetric rotationally/translationally invariant state.
Although $\alpha$ changes in the interval $[-1;1]$ in general, 
the symmetry of the equations with respect to the indexes 
labeling the species reduces the range of $\alpha$  to $[0;1]$. 
Assuming neutron excess implies that the Greek indeces equal 
1 refer to neutrons and while those equal 2 refer to protons.
The relative deformation obviously 
is not bounded and can assume both positive and negative
values. 
\begin{figure}[htb] 
\begin{center}
\includegraphics[angle=-90,width=\linewidth]{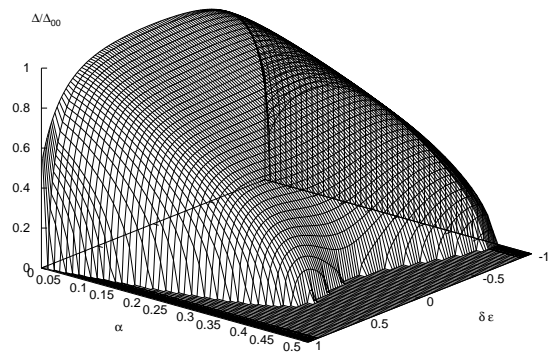}
\end{center}
\caption{The pairing gap as a function of the density asymmetry
  $\alpha$ and the relative deformation $\delta\epsilon$. The gap is 
  normalized to its value for $\delta\alpha =0= \delta\epsilon $.
}
\label{MSfig:fig1}
\end{figure} 
\begin{figure}[htb] 
\begin{center}
\includegraphics[height=3in,width=3.in,angle=-90]{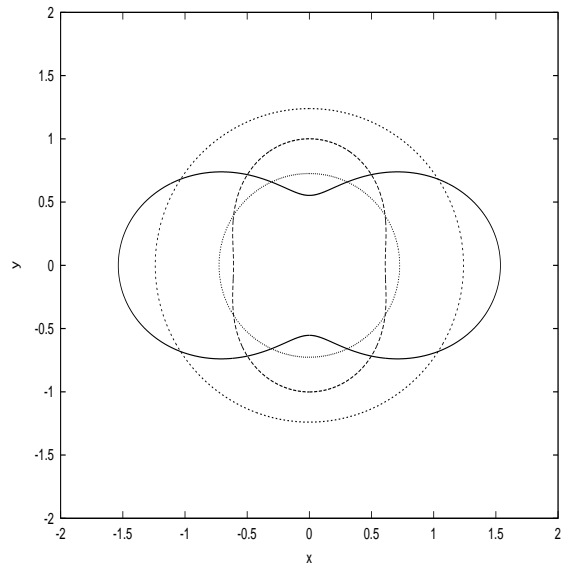}
\end{center}
\caption{A projection of the 
Fermi-surfaces on a plane parallel to the axis of 
the symmetry breaking. The concentric circles correspond to the two 
populations of spin/isospin-up and down fermions in spherically
symmetric state ($\delta\epsilon  = 0$), while the  deformed figures 
correspond to the state with  relative deformation 
$\delta\epsilon = 0.64$. The density asymmetry is $\alpha = 0.35$.
}
\label{MSfig:fig2}
\end{figure} 
For positive values of $\delta\epsilon$, which imply an
oblate deformation for the Fermi-surface of neutrons and a prolate
deformation for the Fermi-surface of protons, the solutions 
for the gap equation show the following features.
For arbitrary constant $\delta\epsilon$ the gap is maximal 
at $\alpha = 0$ and is suppressed as the asymmetry is increased.
For constant $\alpha$, 
$\partial \Delta/\partial \delta\epsilon = 0$ corresponds to 
a maximum at  $\delta\epsilon \neq 0$ in the large $\alpha$ 
limit. The position of the maximum is independent of $\alpha$ 
and is located around $\delta\epsilon = 0.5$ in our model; 
this value also corresponds to the critical asymmetry $\alpha_{c}$
at which the superconducting state vanishes.  Note that 
for $\alpha$ around $\alpha_{c}$ the gap exists only in the deformed
state. For $\alpha = 0$, Eqs. (\ref{GAP3})-(\ref{GR})
are symmetric under exchange of the sign of deformation and so 
is the gap function. In  particular, for $\alpha = 0$, 
the critical deformation for positive and negative deformations coincide. 
For finite $\alpha$ the dependence of the gap on the relative 
deformation depends on the sign of $\delta\epsilon$.
In contrast to the positive range of $\delta\epsilon$, 
where the maximum value of the gap is attained at constant 
$\delta\epsilon$, for negative $\delta\epsilon$ the maximum 
increases as a function of the deformation and saturates around 
$\delta\epsilon \simeq 1$. Quite generally, 
to maintain the maximal phase space overlap, the system prefers 
to keep the sign of  $\delta\mu$ opposite to that of $\delta\epsilon$.
Fig. 2 shows a 2-dimensional projection of a configuration of
deformed Fermi-surfaces for $\delta\epsilon = 0.64$ which
minimizes the free-energy  for fixed $\alpha = 0.35$.
\begin{figure}[htb] 
\begin{center}
\includegraphics[angle=-90,width=\linewidth]{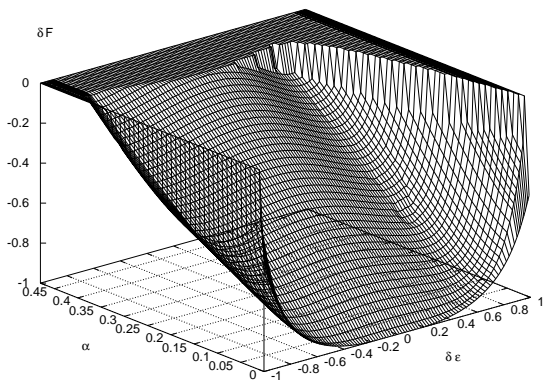}
\end{center}
\caption{The difference in the free-energies of the superconducting
  and normal states $\delta F$ as a function of the density asymmetry
  $\alpha$ and the relative deformation $\delta\epsilon$. The
  free-energy is normalized to its value for $\delta\alpha = 
 \delta\epsilon = 0$.
}
\label{MSfig:fig3}
\end{figure} 

The difference in the free-energies of the superconducting and normal 
states $\delta F$ is shown in Fig. 3. Owing to the symmetries of the
underlying equations, $\delta F$ is symmetric with respect to the sign
change of $\delta\epsilon$ when $\alpha = 0$. For finite $\alpha$'s,
a minor departure from the rotational invariant state leads to 
a decrease in $\delta F$ which develops two minima for either sign 
of $\delta\epsilon$. This behavior can be traced back to the increase 
of the potential energy with increasing gap (cf.
Fig. 1). Note that, although there are non-trivial solutions 
to the gap equation in the large $\alpha$ limit for $\delta\epsilon
\to -1$, these solutions do not lower the energy of the system. The 
superconducting phase becomes unstable for $\alpha > 0.4$ due to the 
increase in the kinetic energy caused by the large deformation of the 
Fermi-surfaces, so that $\delta F$ is a nearly even function 
of $\delta\epsilon$. An inspection of the latent heat
associated with the phase transition at finite temperatures shows that 
this quantity does not vanish at the crossover from the spherically 
symmetric to the deformed superconducting state. Consequently, the 
phase transition associated with the breaking of the rotational
symmetry is of the first order.

To summarize, this paper suggests a novel mechanism of
symmetry breaking in superconducting system with particle-hole 
asymmetry. The lowest order (quadrupole) deformation 
of the Fermi-surfaces (at zero total momentum of the Cooper
pairs) increases the phase-space overlap between the 
Fermi-surfaces of paired quasiparticles, which is otherwise depleted
by the asymmetry in the particle/hole populations. As a result,
the free-energy  develops minima for finite deformations, since
the gain in the (negative) pairing potential energy dominates the 
increase in the kinetic energy caused by the 
deformation. Since the deformed ground state spontaneously breaks 
the rotational symmetry, the dynamic properties of the superconducting
state with deformed Fermi-surfaces such as the sound attenuation, 
the infrared absorption or the Meissner effect will be anisotropic. 
The results above do not depend on the nature of fields 
inducing the asymmetry in the fermion populations nor on the 
nature of the pairing forces and should be applicable to a 
wide range of fermionic systems.


\begin{thebibliography}{99}
\bibitem{COMMENT} The appearance of a two/three-dimensional 
periodic lattice in the LOFF phase breaks, apart from the 
translational symmetry, also the rotational symmetry. 
However, the mechanism of the symmetry breaking
is related to the finite momentum of the Cooper pairs and not to
the deformation of the Fermi-surfaces.
\end{thebibliography}
\end{document}